\documentclass[10pt,letterpaper]{article}

\usepackage{cogsci}

\cogscifinalcopy 

\usepackage{pslatex}
\usepackage{apacite}
\usepackage{float} 

\usepackage{makecell}
\usepackage{array}
\newcolumntype{?}{!{\vrule width 2pt}}

\usepackage{subcaption}
\usepackage{graphicx}
\graphicspath{{figures}}

 \title{Metacognition and Motivation: The Role of Time-Awareness \\ in Preparation for Future Learning}

\author{{ \large \bf Mark Abdelshiheed, Guojing Zhou, Mehak Maniktala, Tiffany Barnes, and Min Chi} \\
  Department of Computer Science\\
  North Carolina State University\\
  Raleigh, NC 27695 \\
  \{{mnabdels,\, gzhou3,\, mmanikt,\, tmbarnes,\, mchi\}}@ncsu.edu}
  
\begin{document}

\maketitle

\begin{abstract}

In this work, we investigate how two factors, \emph{metacognitive skills} and \emph{motivation}, would impact student learning across domains. More specifically, our primary goal is to identify the critical, yet robust, interaction patterns of these two factors that would contribute to students' performance  in learning logic first and then their performance on a \emph{subsequent} new domain, probability. We are concerned with two types of metacognitive skills: \emph{strategy-awareness} and \emph{time-awareness}, that is,  \emph{which} problem-solving strategy to use and \emph{when}  to use it.  Our data were collected from 495 participants across three consecutive semesters, and our results show that the only students who consistently outperform their peers across both domains are those who are not only highly motivated but also strategy-aware and time-aware.  


Intelligent Tutoring Systems
\end{abstract}

\section{Introduction}

Much of prior research has shown that  metacognitive skills,  motivation, or the interaction of the two can have significant impacts on student learning  \cite{efklides2011metacogMotivAndLearning,zimmerman2011metacogMotivAndLearning2}. On the other hand, one fundamental goal of education is to \emph{prepare students for future learning (PFL)} \cite{bransford1999transferRethinking}. While prior research has shown that PFL can be accelerated by obtaining metacognitive skills \cite{zepeda2015TransferANDMotivationOthersthanSelfEfficacy,otieno2013motivationQuestionnairevsOnline,chi2010backward2metacogStrategy,belenky2009metacogTransfer} or influenced by learner's motivation \cite{belenky2013motivationAndTransfer,belenky2012motivationAndTransfer,nokes2011incorporatingMotivationAndTransfer}, no prior work investigated on how the \emph{interactions} of the two could impact PFL. In this work, we investigated how the interactions of the metacognitive skills and motivation  would impact student learning outcomes, as well as  PFL. 

We focused on two types of metacognitive skills related to the students' problem-solving strategies: \emph{strategy-awareness} and \emph{time-awareness}, that is,  \emph{which strategy} to use and \emph{when} to use it. In deductive task domains such as logic and probability,  solving a problem often requires producing an argument, proof, or derivation consisting of one or more inference steps, and each step is the result of applying a domain principle, operator, or rule. Prior work has shown that students often use a mixture of problem-solving strategies such as forward chaining (FC) and backward chaining (BC) during their problem solving \cite{priest1992fAndB,simon1978individualfAndB,newell1972humanfAndB}. Many prior studies investigated the impact of teaching students an explicit problem-solving strategy on their learning outcomes \cite{zepeda2015TransferANDMotivationOthersthanSelfEfficacy,chi2007pyrenees2} or  compared students who were taught different types of  strategies \cite{motivationSelfEfficacy,chi2010backward2metacogStrategy}. In this work, we found that  \emph{time-awareness} should be considered as an independent type  of metacognitive skills apart from problem-solving strategies, and  we investigated: 1) how students' knowledge about which problem-solving strategy to use (strategy-awareness) and when to use it (time-awareness) would impact their learning,  2) how such impacts would change if we factor in student motivation, and  3) how would the robust impact of the interactions between the two types of metacognitive skills and motivation influence students' learning in a new domain. 

We investigate these questions in one type of highly interactive  e-learning environment, Intelligent Tutoring Systems (ITSs) \cite{vanlehn2006behavior}. Our data were collected from training 495 computer science undergraduate  students on a logic tutor first and then a probability tutor at North Carolina State University across three consecutive semesters. Both tutors are assigned to students as regular homework assignments: the logic tutor is assigned at the beginning of the semester, while the probability tutor is at the end of the semester. In the logic tutor, students solve problems by applying logic rules to derive new logical statements until reaching the conclusion, by either following a FC-like strategy (the default) or switching to a BC-like strategy. In the probability tutor, students are required to follow a BC strategy to solve the training problems. 

Both tutors have been extensively evaluated in the past decade and a series of papers have been published to show their effectiveness independently \cite{chi2010backward2metacogStrategy,barnes2008dt,chi2007pyrenees1, chi2007pyrenees2,croy2000dt}. This is our first attempt to examine student learning across the two tutors and our goal is to  determine how the interactions between students' motivation and their spontaneous switch of problem-solving strategies in the logic tutor would impact their learning outcomes, and how such interaction patterns would impact their subsequent learning on probability. Overall, our results show that  strategy-awareness and time-awareness would help students learn better in the logic tutor; however, it is only when both types of metacognitive skills are aligned with high motivation, that student learning is significantly improved  in the probability tutor.

\section{Background}
\citeA{bransford1999transferRethinking} proposed the theory of Preparation for Future Learning (PFL) that assumes that students need to continue to learn, and investigates whether students are prepared to do so. Similar to prior work \cite{fancsali2014motivationOnline,otieno2013motivationQuestionnairevsOnline,chi2010backward2metacogStrategy}, we bring PFL into the ITS context, where it is possible to directly measure processes associated with PFL. In this work,  we evaluate students' choices of whether, what, and when to select a specific problem-solving strategy. Based on \citeA{winne2014metacognitionHeaven},  mastering strategy selection alone is a cognitive skill, but when incorporated with awareness about when such strategy should be changed, the skill becomes a metacognitive skill. Therefore, we argue that \emph{strategy-awareness} and \emph{time-awareness} should be two different types of metacognitive skills. In this work, we investigated how the \emph{interactions} of the two types of metacognitive skills and motivation could impact PFL. Next, we review related work. 





\subsection{The Impact of Metacognition on Transfer}

Metacognition indicates one's realization of their own cognition as well as being able to regulate it \cite{chambres2002metacognitionStrategySelection,roberts1993metacogDefinitionStrategySelection2}. It is the act of exercising and monitoring control of cognitive skills \cite{efklides2011metacogMotivAndLearning}. Hence, a metacognitive skill often consists of a cognitive skill and a regulator for controlling this skill.

Many studies have shown that  metacognitive skills have positive impacts on learning  \cite{zepeda2019metacogLearning,zepeda2015TransferANDMotivationOthersthanSelfEfficacy}, on transfer across ITSs \cite{zepeda2015TransferANDMotivationOthersthanSelfEfficacy,chi2010backward2metacogStrategy}, as well as on students' learning behaviors \cite{belenky2009metacogTransfer}. Several approaches have been used to evaluate metacognitive skills, such as tutoring prompts \cite{zepeda2015TransferANDMotivationOthersthanSelfEfficacy,belenky2009metacogTransfer}, strategy selection \cite{chi2010backward2metacogStrategy,roberts1993metacogDefinitionStrategySelection2}, and reading comprehension and memory recall \cite{chambres2002metacognitionStrategySelection}. 

\citeA{zepeda2015TransferANDMotivationOthersthanSelfEfficacy} demonstrated that metacognitive instruction could influence student metacognitive skills, motivation, and transfer learning. Students who were taught planning, monitoring, and evaluating made better metacognitive judgments and demonstrated higher motivation levels (e.g., task value and self-efficacy) than those who were not. As an example of PFL, the former also performed better on a novel self-guided learning task than the latter. In our prior work, \citeA{chi2010backward2metacogStrategy} investigated the transfer of metacognitive skills from a probability tutor to a physics tutor. We
showed that an ITS teaching domain-independent metacognitive skills could close the gap between high and low learners, not only in the domain where they were taught (probability), but also in a second domain where they were not taught (physics). In that study, the metacognitive skills included a  problem-solving strategy and principle-emphasis instructions. We found that it was the principle-emphasis skill that is transferred across the two domains and that closed the gap between the high and low learners. 

In summary, most aforementioned studies showed that explicitly teaching students metacognitive skills such as problem-solving strategies can improve their learning outcomes not only in the domain they were taught but also in a new domain. In this work, we investigated how  \emph{students' own} metacognitive skills  including  strategy-awareness and time-awareness would impact their learning and also prepare them for future learning in a new domain with a new ITS.      


\subsection{The Impact of Motivation on Learning and Transfer}


Substantial work has shown that motivation can significantly impact learning  \cite{motivationSelfEfficacy,bernacki2014motivationStability,fancsali2014motivationOnline,bernacki2013motivationChange}. For example,  \citeA{motivationSelfEfficacy} investigated how self-efficacy could  impact students'  procedural or conceptual knowledge when learning physics. They showed that self-efficacy was related to only conceptual knowledge, which implies that highly motivated students tend to understand, rather than memorize, principles. \citeA{bernacki2014motivationStability} showed that students' motivation may change over time, and such changes may impact their learning outcomes in a geometry tutor. They found that students with stable mastery-approach goals (aim to master the task)  achieved higher grades than those with variable ones. Additionally, students with stable performance-approach goals (aim to outperform others) requested fewer hints than those with variable ones.

Many studies have also shown the impact of motivation on transfer \cite{belenky2013motivationAndTransfer,belenky2012motivationAndTransfer,nokes2011incorporatingMotivationAndTransfer}. For instance, \citeA{belenky2012motivationAndTransfer} found that PFL is influenced by the interaction of students’ motivation (achievement goals) and different forms of instruction. Students who had high mastery-approach goal orientation showed signs of transfer, irrespective of the instruction type. Moreover, students who were allowed to innovate new strategies developed higher motivation aspects, compared to those who followed direct instruction. Later, innovative students showed higher signs of transfer when given a final wording problem. \citeA{nokes2011incorporatingMotivationAndTransfer} incorporated students' achievement goals into a framework that accounts for transfer success or failure. The framework represents a loop of goal generation, environment interpretation, knowledge \& goal representation, solution generation, and solution evaluation. The last step decides whether the loop will be repeated or not. They tested this framework, and reported that mastery-approach goal-oriented students were more likely to succeed in knowledge transfer.


One of the crucial questions for research on motivation is how to define and measure motivation.  \citeA{eccles1983motivationexpectancy} defined motivation to be a process that combines the individuals' perception of three factors: expectations for success, subjective task value, and intrinsic interest. \citeA{toure2014motivationDimension} extended this definition and classified motivation into two dimensions: \textit{outcome-focused} `getting it done' and \textit{process-focused} `doing it right'. To measure motivation, prior research  explored self-efficacy \cite{kalender2019motivationSurvey2,motivationSelfEfficacy,zepeda2015TransferANDMotivationOthersthanSelfEfficacy,fancsali2014motivationOnline,bernacki2013motivationChange}, goal orientation \cite{otieno2013motivationQuestionnairevsOnline,belenky2013motivationAndTransfer,belenky2012motivationAndTransfer}, sense of belonging \cite{kalender2019motivationSurvey2} and accuracy \cite{toure2014motivationDimension}. The majority of these studies used surveys to measure these aspects. For instance, \citeA{kalender2019motivationSurvey2} used a survey to measure three motivational aspects based on the achievement goals: self-efficacy, interest, and sense of belonging. \citeA{motivationSelfEfficacy} utilized a survey to measure self-efficacy, in addition to three achievement goals: mastery-approach, performance-approach and performance-avoidance. In recent years, digital technologies such as ITSs made it possible to measure motivation using students' online traces  \cite{fancsali2014motivationOnline,bernacki2013motivationChange,otieno2013motivationQuestionnairevsOnline}. For instance,
\citeA{otieno2013motivationQuestionnairevsOnline} used the online use of hint and glossaries in a geometry tutor as a motivation measure and found that the online measures differ from the motivation measures using questionnaire data, and the former was more predictive of posttest scores than the latter. Therefore, in this work, we also used students' online traces to measure their motivation levels. 



\section{Experiment}
Our data were collected from an undergraduate computer science course at North Carolina State University across three  semesters. Students were trained on the logic tutor first, and then on the probability tutor at least two months later.  A total of $495$ students finished both tutors: $N=151$ for Fall 2017, $N=128$ for Spring 2018, and $N=216$ for Fall 2018. 

\subsection{Methods}

\subsubsection{Our Logic and Probability tutors:}  All students went through a standard pretest-training-posttest procedure on each tutor.

\begin{figure}[h]
     \centering
     \begin{subfigure}[t]{0.23\textwidth}
         \centering
         \includegraphics[width=\textwidth]{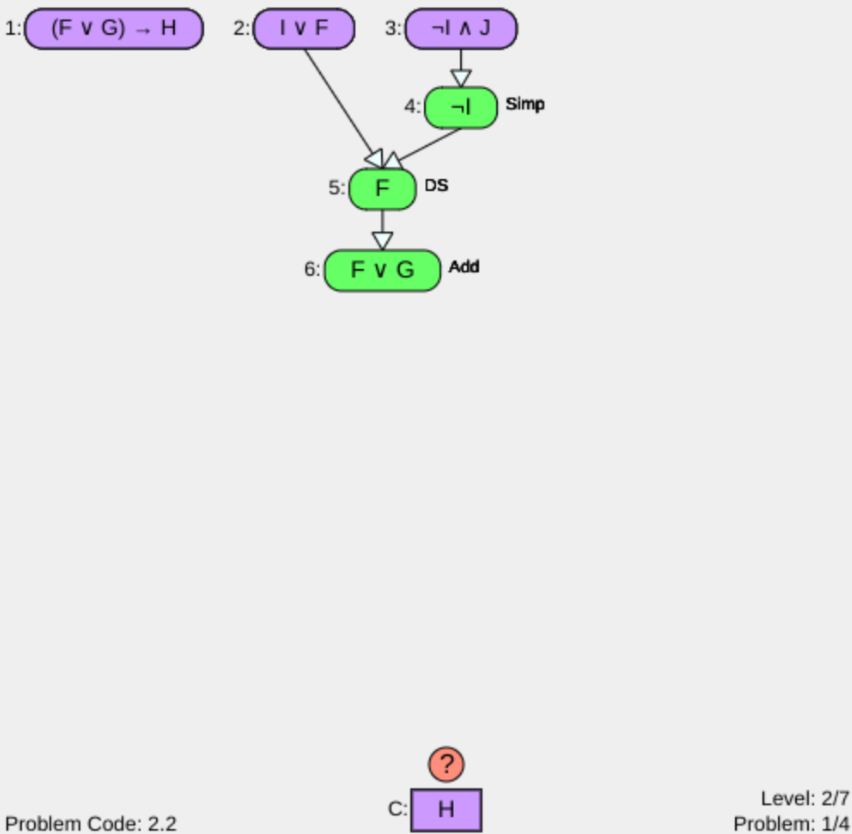}
         \caption{Direct Proof}
         \label{fig:direct}
     \end{subfigure}
     \hfill
     \begin{subfigure}[t]{0.23\textwidth}
         \centering
         \includegraphics[width=\textwidth, ]{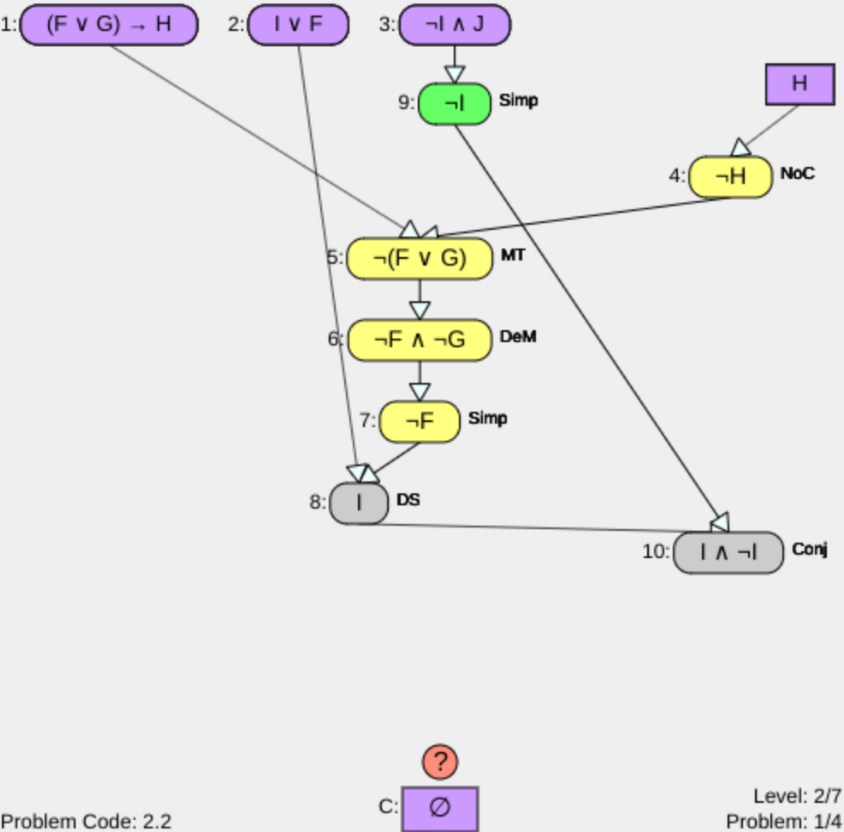}
         \caption{Indirect Proof}
         \label{fig:indirect}
     \end{subfigure}
\caption{Logic Tutor Problem-Solving Strategies}
\label{DT}
\end{figure}

 Our logic tutor teaches students how to prove propositional logic statements. A student can solve a problem in one of two strategies: \textbf{direct} or \textbf{indirect}. Figure \ref{fig:direct} shows that in \emph{direct} proofs,  a student needs to derive the conclusion node at the bottom from the givens at the top;  while Figure \ref{fig:indirect} shows that in \emph{indirect} proofs, a student needs to derive a contradiction from the givens and the \emph{negation} of the conclusion. Both logic pre- and  posttests  have two problems and their scores are functions of time and accuracy.  The training on the logic tutor includes around $20$ problems.

\begin{figure}[ht]
\begin{center}
\includegraphics[width=8cm,height=5cm]{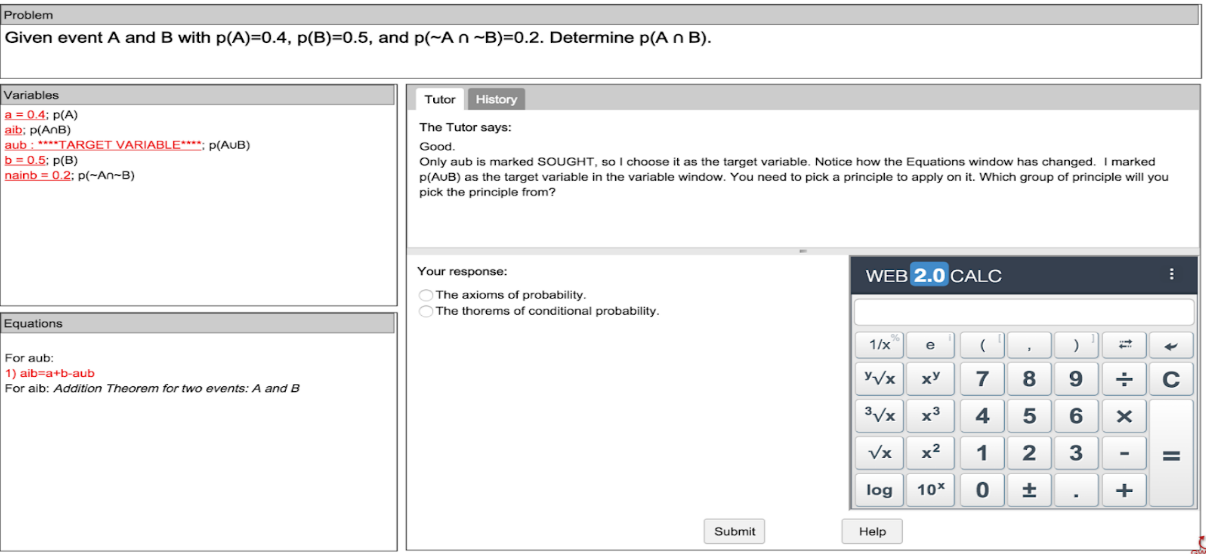}
\end{center}
\vskip-0.12in
\caption{Probability Tutor Interface} 
\label{fig:pyr}
\end{figure}

Figure \ref{fig:pyr} shows the GUI for our probability tutor that teaches students how to apply principles to solve probability problems. The pre- and posttest sections have $14$ and $20$ problems, respectively. These problems require students to derive an answer by writing and solving one or more equations.  The training includes $12$ problems.  

\noindent There are \textbf{two major differences} between the two ITSs: 

1. In the probability tutor, our pre- and posttest scores are based on accuracy. Both tests were graded in a double-blind manner by experienced graders. In the logic tutor, they are based on both accuracy and efficiency. Since there are only two questions in each test, the class instructor believes that it is as important for students to solve them accurately as for them to solve them quickly.  For comparison purposes, all test scores were normalized to the range of $[0,100]$. Note that in both tutors, the posttest is \emph{much harder} than the pretest.

2. In the logic tutor, students can select FC-like direct proofs (the default), or choose to switch to BC-like indirect proofs. Conversely, in the probability tutor, students can only use BC. In both tutors,  students can employ any strategy during the pre- and posttest.

\subsubsection{Metacognitive Skills}

Students can choose to switch problem-solving strategies \emph{only} when training on the logic tutor. Thus, we measured students' metacognitive skills based on their interactions with the logic tutor alone. The training problems in the logic tutor are organized into \textbf{five}  ordered levels with an \emph{incremental degree of difficulty} and in each level, students are required to complete four problems. Each problem can be solved by either following the default FC or switching to BC. However, most problems especially the higher level ones can be solved much more efficiently by BC, and we expect that effective problem solvers should switch their strategy on these problems, and more importantly,  they should switch it early when solving them. Thus, our metacognitive skill measurement is a combination of \emph{strategy-awareness}:  using the default direct proof or switching to indirect proof \cite{chi2010backward2metacogStrategy,chambres2002metacognitionStrategySelection,roberts1993metacogDefinitionStrategySelection2}, and \emph{time-awareness}: when such switch happens  \cite{winne2014metacognitionHeaven}. We considered two factors in time-awareness: one is that a student should switch in later levels (harder training problems) where the savings will be big and the other is that students should switch early (when convenient) during solving a problem. On average,  students take 210 actions to solve a problem, and the median number of actions that a student took before switching is 30.  Therefore, we calculated metacognitive score (MetaScore) for a student $i$ as:

$MetaScore_i = \sum_{L=1}^{5} [\sum_{p=1}^{4} [L* SAware_{ip} * TAware_{ip} ]]$

\noindent where strategy-awareness $SAware_{ip} = 1$ indicates that student $i$ switched strategy when solving problem $p$ at level $L$, while $0$ means no switch; for time-awareness $TAware_{ip} = 1$  when the student $i$ switched early on problem $p$ ($\le 30$ actions) and $TAware_{ip} =-1$ for late switch ($> 30$ actions). Based on this formula,  $MetaScore_i >0$  indicates that student $i$ is both strategy-aware and time-aware; if $MetaScore_{i} <0$, it indicates that student $i$ is strategy-aware but not good at knowing when to switch (time-unaware); finally, if $MetaScore_{i} =0$, we do not have enough evidence on the student's metacognitive skills in that he/she may simply follow the default FC settings. Based on  $MetaScore$s, students are divided into three groups: those who showed both strategy- and time-awareness  ($MetaScore >0$) are referred to as the  `Str\_Time' group  $(N = 145)$; those who showed  strategy awareness only  ($MetaScore <0$) as `Str Only' $(N = 166)$; and  the default students ($MetaScore =0$) as  `Default' $(N = 184)$. 

\subsubsection{Motivation}
Motivated by prior research on motivation \cite{toure2014motivationDimension},  we measured students' motivation by tracking the accuracy of their online traces. By doing so, we factor in the fact that students often have different motivations: some are more outcome-oriented for better grades and some are more process-focused for learning the domain subject as much as possible.  Similar to prior work \cite{earlymotivation2,earlymotivation1}, students' motivation in this work is defined based on their initial interactions in the early stages of \emph{each} tutor. More specifically, we used the percentage of correct rule applications in the first two problem-solving questions as our motivation indicators. In other words, our measured students' \emph{initial} motivation levels do not consider the fact that students' motivation levels may change over time. Students are divided into high- and low-motivation groups through a median split. For logic: $HM_{Logic}$ $(N = 248)$ and $LM_{Logic}$ $(N = 247)$ and for probability: $HM_{Prob}$ $(N = 249)$ and  $LM_{Prob}$ $(N = 246)$. A chi-square test shows no significant evidence on students staying in the same motivation level across the two tutors: $\chi^2 (1,\, N=495) = 1.26, \, p=0.26$. In other words, students' motivation levels may change over a course of a semester or change based on the subjective domains. Additionally, our motivation definition differs from students' incoming competence in that one-way ANOVA showed no significant difference in the pretest score between the high- and low-motivation students: $\mathit{F}(1,493) = 0.7,\, \mathit{p} = .17$ for Logic and $\mathit{F}(1,493) = 0.001,\, \mathit{p} = .98$ for Probability.

\section{Results}

We will examine  the impact of 1)  metacognitive skills alone, 2) motivation alone, and 3) the interactions of the two on students' learning across both tutors. For each tutor, students' learning performance is measured using their corresponding pre- and posttest scores, together with their normalized learning gain (NLG) defined as: $(NLG = \frac{post - pre}{\sqrt{100 - pre}})$, where 100 is the maximum posttest score. For reporting convenience, pre- and  posttest scores are normalized to the range of $[0,100]$.

 

\subsection{Metacognitive Skills}

\begin{table}[h!]
\begin{center} 
\caption{Comparing the three Metacognitive Groups} 
\label{MetaPostNLG} 
\begin{tabular}{lllll} 
\Xhline{4\arrayrulewidth}
\multicolumn{5}{c}{Logic Tutor}\\
\hline
Group   & Size & Pre & Post & NLG \\
\hline
Str\_Time & 145   & 78.4 (3.2) & 75.8 (1.7) & 0.94 (.395) \\
Str Only & 166 & 74.9 (3)  & 68.2 (1.67) &  -0.46 (.39) \\
Default & 184  & 75.5 (2.8) & 70.9 (1.68)  & 0.19 (.393) \\
\hline
\multicolumn{5}{c}{Probability Tutor}\\
\hline
Group   & Size & Pre & Post & NLG \\
\hline
Str\_Time & 145   & 72.3 (2.8) & 75.5 (3) & 0.02 (.06) \\
Str Only & 166 & 72.1 (2.5)  & 74 (2.8) &  0.01 (.05) \\
Default & 184  & 71.8 (2.6) & 73.4 (2.6)  & -0.007 (.05) \\
\Xhline{4\arrayrulewidth}
\end{tabular} 
\end{center} 
\end{table}


Table \ref{MetaPostNLG} above compares the three metacognitive groups' learning performances on the logic tutor (upper) and probability tutors (lower) respectively. It shows  the mean and SD  of the pretest scores (Pre), the overall posttest scores (Post), and the NLGs. For the logic tutor, while no significant difference was found among the three groups on Pre, a one-way ANCOVA analysis on metacognitive groups using the pretest score as a covariate showed a significant difference in their posttest scores: ($\mathit{F}(2,491) = 17.3,\, \mathit{p} < .001, \, \mathit{\eta} = 0.3$). Subsequent contrast analyses showed that `Str\_Time' scored significantly higher than both `Str Only': $\mathit{t}(309) = 5.8,\, \mathit{p} <.0001,\, \mathit{d} = 4.5$ and `Default': $\mathit{t}(327) = 3.8,\, \mathit{p} < .001,\, \mathit{d} = 2.9$. Additionally, `Default' scored significantly higher than `Str Only': $\mathit{t}(348) = 2.2,\, \mathit{p} = .03,\, \mathit{d} = 1.6$. While a one-way ANOVA showed no significant difference among the three groups on the NLGs,  subsequent contrast analyses showed that `Str\_Time' scored significantly higher than `Str Only': $\mathit{t}(309) = 2.4,\, \mathit{p} = .02,\, \mathit{d} = 3.6$. For the probability tutor, however, no significant results were found among the three metacognitive groups on either Pre-, Posttest, or NLGs. 

To summarize,  our results suggest that strategy-awareness alone cannot lead students to learn better in logic; students need to be time-aware as well. Additionally, while `Str\_Time' group learns significantly  better than the other two groups in logic, they did not outperform other groups in probability.



\subsection{Motivation Level}


Table \ref{MotivPostNLG}  compares the high- and low-motivation groups' learning performances on the logic and probability tutors respectively. As mentioned before, no significant difference was found between the high- and low-motivation groups on the pretest on either tutor. As expected, a one-way ANCOVA analysis using motivation as a factor and pretest as a covariate showed that on both tutors, high-motivation students scored significantly higher than their low peers on the corresponding posttest: $\mathit{F}(1,492) = 15.8,\, \mathit{p} < .001, \, \mathit{\eta} = 0.17$ for the logic tutor  and $\mathit{F}(1,492) = 24.5,\, \mathit{p} < .001, \, \mathit{\eta} = 0.17$ for the probability tutor. While no significant difference was found between the two groups' NLGs in our logic tutor, one-way ANOVA showed that highly motivated students had significantly higher NLGs than the low ones in the probability tutor: $\mathit{F}(1,493) = 7.6,\, \mathit{p} < .01, \, \mathit{\eta} = 0.12$.  In short, our results suggest that our motivation measure is reasonable in that:   the highly motivated students indeed performed significantly better than the low ones on the posttest on both the logic and probability tutors. The former also had significantly higher NLG than the latter on  the probability tutor.

\begin{table}[H]
\begin{center} 
\caption{Comparing the motivation level in each tutor} 
\label{MotivPostNLG} 
\begin{tabular}{lllll} 
\Xhline{4\arrayrulewidth}
\multicolumn{5}{c}{Logic Tutor}\\
\hline
Group   & Size & Pre & Post & NLG \\
\hline
$HM_{Logic}$ & 248   & 78.9 (5.3) & 73.6 (1.4) & 0.25 (.06) \\
$LM_{Logic}$ & 247 & 73.4 (5.5)  & 69.2 (1.4) &  0.14 (.07) \\
\hline
\multicolumn{5}{c}{Probability Tutor}\\
\hline
Group    &  Size  & Pre & Post & NLG \\
\hline
$HM_{Prob}$   &  249 & 81.7 (4.2) &79 (1.8)  & 0.05 (.04)  \\
$LM_{Prob}$   &   246 & 77 (4.4) & 69 (2.5) & -0.03 (.04) \\
\Xhline{4\arrayrulewidth}
\end{tabular} 
\end{center} 
\end{table}

\subsection{Interaction Between Metacognition and Motivation }

\subsubsection{Logic Tutor:}
Combining the three metacognitive groups: `Str\_Time', `Str Only', and `Default' with the two motivation levels: $HM_{Logic}$ and $LM_{Logic}$ resulted in six groups. A chi-square test showed students' motivation level did not differ significantly across the three metacognitive groups: $\chi^2 (2,\, N=495) = 2.87, \, p=0.24$. Additionally, no significant difference was found among the six groups on the logic pretest: $\mathit{F}(2,489) = 0.69,\, \mathit{p} = .49$. 

\begin{figure}[b!]
\begin{center}
\includegraphics[width=8cm,height=5cm]{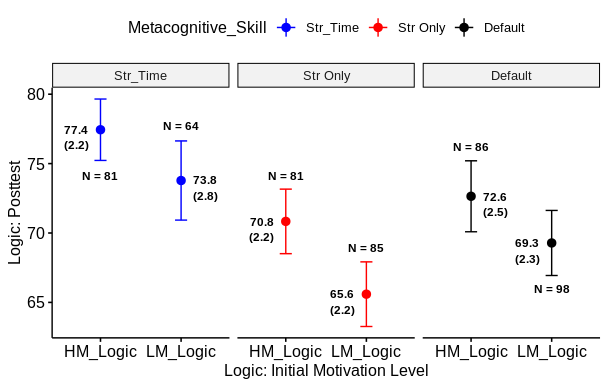}
\end{center}
\caption{Logic: Metacognition \& Motivation vs Posttest} 
\label{fig:crossDTPost}
\end{figure}

Figure \ref{fig:crossDTPost} compares the six groups' performance on the logic posttest.  A two-way ANCOVA using the metacognitive groups  and motivation levels as factors and pretest as covariate showed no significant interaction effect. However, there was a  main effect of metacognitive groups: $\mathit{F}(2,488) = 16.6,\, \mathit{p} < .0001$, and a main effect of motivation level: $\mathit{F}(1,488) = 16.7,\, \mathit{p} < .0001$. More specifically, within each metacognitive group, the  $HM_{Logic}$ group significantly outperformed the corresponding $LM_{Logic}$ group: $\mathit{t}(143) = 2,\, \mathit{p} = .04,\, \mathit{d} = 1.4$ for  `Str\_Time', $\mathit{t}(164) = 3.1,\, \mathit{p} < .01,\, \mathit{d} = 2.4$ for `Str Only' and $\mathit{t}(182) = 2.1,\, \mathit{p} = .03,\, \mathit{d} = 1.4$ for `Default'. Among the three $HM_{Logic}$ groups, the high-motivation `Str\_Time' students scored significantly higher than their peers: $\mathit{t}(160) = 3.8,\, \mathit{p} < .001,\, \mathit{d} = 3$ against the high-motivation `Str Only' peers and  $\mathit{t}(165) = 2.8,\, \mathit{p} < .01,\, \mathit{d} = 2.1$ against the high-motivation `Default' ones.  Among the three $LM_{Logic}$ groups, the low-motivation `Str\_Time' group scored significantly higher than the low-motivation `Default' group:  $\mathit{t}(160) = 2.5,\, \mathit{p} = .02,\, \mathit{d} = 1.8$ and the latter scored significantly higher than the  low-motivation `Str Only' group: $\mathit{t}(181) = 2.2,\, \mathit{p} =.03,\, \mathit{d} = 1.6$. 

Similarly, a two-way ANOVA using the two factors found no significant interaction effect on NLGs nor any main effect. However, among the three $HM_{Logic}$ groups, the high-motivation `Str\_Time' group scored significantly higher than both the high-motivation `Str Only' group: $\mathit{t}(160) = 2.3,\, \mathit{p} = .03,\, \mathit{d} = 5.4$ and the high-motivation `Default' group:  $\mathit{t}(165) = 2.2,\, \mathit{p} = .03,\, \mathit{d} = 3.5$.   No significant difference was found among the three $LM_{Logic}$ groups. Additionally, only within the two `Str\_Time' groups, the high-motivation `Str\_Time' students scored significantly higher than their low-motivation `Str\_Time' peers.  No significant difference between the high- and low-motivation groups was found within `Default' and `Str Only'.
In short, our results suggest that  the high-motivation `Str\_Time' group performs the best among the six groups in terms of both posttest scores and NLGs on the logic tutor.






\subsubsection{Probability Tutor: } 
\begin{figure}[b!]
\begin{center}
\includegraphics[width=8cm,height=5cm]{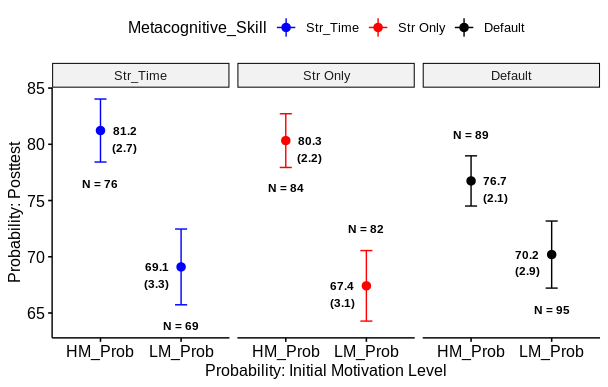}
\end{center}
\caption{Metacog (Log) X Motiv (Prob): Posttest (Prob)} 
\label{fig:crossPYRPost}
\end{figure}

\begin{figure}[h!]
\begin{center}
\includegraphics[width=8cm,height=5cm]{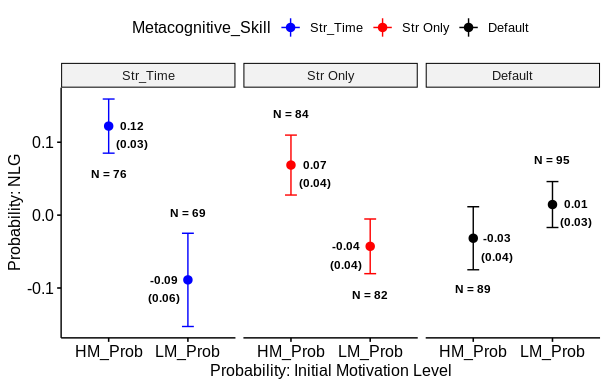}
\end{center}
\caption{Metacog (Log) X Motiv (Prob): NLG (Prob)} 
\label{fig:crossPYRNLG}
\end{figure}

Similarly, we combined the three metacognitive groups  with the two motivation levels defined based on students' interactions on the probability tutor: $HM_{Prob}$ and $LM_{Prob}$, resulting in six groups. A chi-square test showed students' motivation level on the probability tutor did not differ significantly across the three metacognitive groups: $\chi^2 (2,\, N=495) = 0.53, \, p=0.76$. Moreover, no significant difference was found among the six groups on the probability pretest: $\mathit{F}(2,489) = 0.5,\, \mathit{p} = .63$. Figures \ref{fig:crossPYRPost} and \ref{fig:crossPYRNLG} compare the six groups' probability posttest and NLGs respectively.

A two-way ANCOVA with metacognitive skills and motivation as factors and  pretest scores as covariate, showed a significant interaction effect on the posttest: $\mathit{F}(2,488) = 3.8,\, \mathit{p} = .02, \, \mathit{\eta} = 0.09$. Additionally, there was a  main effect of motivation in that high-motivation students scored significantly higher than their low peers: $\mathit{F}(1,488) = 24.4,\, \mathit{p} < .0001$. Among the three highly motivated groups, both `Str\_Time' and `Str Only' scored significantly higher than `Default': $\mathit{t}(163) = 2.4,\, \mathit{p} = .02,\, \mathit{d} = 1.9$ and $\mathit{t}(171) = 2.4,\, \mathit{p} = .02,\, \mathit{d} = 1.7$, respectively. However, no such difference was found among the three low-motivation groups.

 Similarly, as shown in Fig  \ref{fig:crossPYRNLG}, a two-way ANOVA using metacognitive skills and motivation as factors showed a significant interaction effect on NLGs: $\mathit{F}(2,489) = 6.4,\, \mathit{p} < .01, \, \mathit{\eta} = 0.16$ and also there was a  main effect of motivation: $\mathit{F}(1,489) = 7.8,\, \mathit{p} < .01$. Subsequent contrast analyses showed that: high-motivation `Str\_Time' students scored significantly higher than their low peers: $\mathit{t}(143) = 3.8,\, \mathit{p} < .001,\, \mathit{d} = 4.4$. The same pattern was observed between the two `Str Only' groups: $\mathit{t}(164) = 2.2,\, \mathit{p} = .03,\, \mathit{d} = 2.9$. Across the three high-motivation groups,  both `Str\_Time' and `Str Only' scored significantly higher than their `Default' peers: $\mathit{t}(163) = 3,\, \mathit{p} < .01,\, \mathit{d} = 4.2$ and $\mathit{t}(171) = 2,\, \mathit{p} = .04,\, \mathit{d} = 2.5$, respectively. In short, on our probability tutor,  the high-motivation `Str\_Time' group performs the best  among the six groups, on both posttest scores and NLGs.

\section{Conclusion}

In this work,  we investigate how two factors, \emph{metacognitive skills} and \emph{motivation}, would impact student learning across two domains: first logic and then probability.  Our results from  analyzing 495 students' performance on two tutors show that when considering each factor alone, no consistent robust pattern is found, while when combining the two factors, we find that the students who are \emph{highly motivated}, \emph{strategy-aware}, and \emph{time-aware} consistently outperform their peers across both domains.

Firstly and most importantly, our analysis results confirm the importance of motivation in that across both tutors, the impacts of metacognitive skills on student learning only appear among the highly motivated student groups and for low motivated students, no significant difference was found among the three metacognitive skills groups. In other words, our results reveal an aptitude-treatment interaction (ATI) effect \cite{kanfer1989ati} in that some students may be insensitive to learning unless the presented material matches their aptitude.  While such findings are not surprising, they show that it is crucial for further research  to understand why certain students are not motivated, and to explore how to motivate them. Moreover, our findings confirm that our choice of using students’ online traces on the first two questions is a reasonable way to measure their motivation levels. 

Secondly, while problem-solving strategies have been extensively explored in prior research, as far as we know this is the first work that investigates students' metacognitive skills from both  \emph{strategy-aware} and \emph{time-aware} aspects. Our results suggest that these two skills are indeed different in that while both `Str\_Time' and `Str Only' groups know about problem-solving strategies, only the former group knows \emph{when} to apply them. More importantly, it is essential to consider the \emph{time-aware} aspect when assessing students' metacognitive skills in that when highly motivated, `Str\_Time' consistently outperforms `Str Only' and `Default' on both tutors.  

Thirdly, our results show that `Str Only' can benefit greatly by training on an ITS that explicitly teaches and follows problem-solving strategies. While the high-motivation `Str Only' performed the worst than their `Str\_Time' and `Default' peers on the logic tutor, they performed as well as the high-motivation `Str\_Time' and both outperformed their `Default' peers on the probability tutor. One potential explanation is that the \emph{time-aware} aspect of the skills is not needed when training on our probability tutor, since it follows the same explicit problem-solving strategy on all problems.

Finally, we emphasize the importance of mastering different problem-solving strategies for highly motivated students, and its role on PFL. We found that only across the highly motivated groups, both `Str Only' and `Str\_Time'  had significantly higher probability scores than the `Default' group. This finding suggests evidence for metacognitive skill transfer in highly motivated students who are aware of switching strategies. In other words, while time awareness could be a decisive factor for consistency, strategy awareness might identify students who are prepared for future learning.

Despite these findings, it is important to note that there is at least one caveat in our analysis: we measured students' motivation using the first two problems on each tutor and we did not consider that students' motivation levels may vary during the training. Furthermore, as for  future work,  we will investigate whether explicitly prompting students to switch their problem-solving strategy would boost `Str Only' and  `Default' students' performance  so that they can catch up with the `Str\_Time' students.

\noindent \textbf{Acknowledgments:} This research was supported by the NSF Grants: 2013502, 1726550,  1660878 and 1651909.

\bibliographystyle{apacite}

\setlength{\bibleftmargin}{.125in}
\setlength{\bibindent}{-\bibleftmargin}

\bibliography{cogsci2020}

\end{document}